\documentclass[snbasic]{sn-jnl}% Basic Springer Nature Reference Style/Chemistry Reference Style
\usepackage{natbib} 
\usepackage{graphicx}%
\usepackage{multirow}%
\usepackage{amsmath,amssymb,amsfonts}%
\usepackage{amsthm}%
\usepackage{mathrsfs}%
\usepackage[title]{appendix}%
\usepackage{xcolor}%
\usepackage{textcomp}%
\usepackage{manyfoot}%
\usepackage{booktabs}%
\usepackage{algorithm}%
\usepackage{algorithmicx}%
\usepackage{algpseudocode}%
\usepackage{listings}%

\usepackage{makecell}
\usepackage[T1]{fontenc}
\usepackage[utf8]{inputenc}
\usepackage{braket}
\usepackage{graphicx,epstopdf}
\usepackage{bm}
\usepackage{mathtools}
\usepackage{amsmath}
\usepackage{amssymb}
\usepackage{mathtools}
\usepackage{amsthm}

\DeclareGraphicsExtensions{.png.pdf}

\newcommand{\bfm}[1]{\ensuremath{\mathbf{#1}}}

\def \bstheta{\bfsym \theta}

\def\bx{\bfm x}

 \def\cC{{\cal  C}}

\def\a\cos{\mathrm{arc\cos}}

\newcommand{\bfsym}[1]{\ensuremath{\boldsymbol{#1}}}

\def\bstheta{\bfsym {\theta}}

\def\bstheta{\bfsym \theta}

\begin{document}

  \title[Classifying Metamorphic versus Single-Fold Proteins with Statistical Learning and AlphaFold2]{Classifying Metamorphic versus Single-Fold Proteins with Statistical Learning and AlphaFold2}

\author[1]{\fnm{Yongkai} \sur{Chen}}%\email{yongkaichen@fas.harvard.edu}

\author*[2]{\fnm{Samuel W.K.} \sur{Wong}}\email{samuel.wong@uwaterloo.ca}

\author*[1]{\fnm{S. C.} \sur{Kou}}\email{kou@stat.harvard.edu}

\affil[1]{\orgdiv{Department of Statistics}, \orgname{Harvard University}, \orgaddress{\street{1 Oxford Street}, \city{Cambridge}, \postcode{02138}, \state{MA}, \country{United States}}}
\affil[2]{\orgdiv{Department of Statistics and Actuarial Science}, \orgname{University of Waterloo}, \orgaddress{\street{200 University Avenue West}, \city{Waterloo}, \postcode{N2L 3G1}, \state{ON}, \country{Canada}}}

\abstract{The remarkable success of AlphaFold2 in providing accurate atomic-level prediction of protein structures from their amino acid sequence has transformed approaches to the protein folding problem. However, its core paradigm of mapping one sequence to one structure may only be appropriate for single-fold proteins with one stable conformation. Metamorphic proteins, which can adopt multiple distinct conformations, have conformational diversity that cannot be adequately modeled by AlphaFold2. Hence, classifying whether a given protein is metamorphic or single-fold remains a critical challenge for both laboratory experiments and computational methods. To address this challenge, we developed a novel classification framework by re-purposing AlphaFold2 to generate conformational ensembles via a multiple sequence alignment sampling method. From these ensembles, we extract a comprehensive set of features characterizing the conformational ensemble's modality and structural dispersion. A random forest classifier trained on a carefully curated benchmark dataset of known metamorphic and single-fold proteins achieves a mean AUC of 0.869 with cross-validation, demonstrating the effectiveness of our integrated approach. Furthermore, by applying our classifier to 600 randomly sampled proteins from the Protein Data Bank, we identified several potential metamorphic protein candidates -- including the 40S ribosomal protein S30, whose conformational change is crucial for its secondary function in antimicrobial defense. By combining AI-driven protein structure prediction with statistical learning, our work provides a powerful new approach for discovering metamorphic proteins and deepens our understanding of their role in their molecular function.}

  \maketitle

%\end{frontmatter}
%%%%%%%%%%%%%%%%%%%%%%%%%%%%%%%%%%%%%%%%%%%%%%
%% Please use \tableofcontents for articles %%
%% with 50 pages and more                   %%
%%%%%%%%%%%%%%%%%%%%%%%%%%%%%%%%%%%%%%%%%%%%%%
%\tableofcontents
\section{Introduction}\label{sec:intro}
Proteins were famously called the ``machines of life'' by Max Perutz, the Nobel laureate who first discovered the three-dimensional (3D) structure of hemoglobin using X-ray methods \citep{perutz1960structure}. The vital roles of proteins in living organisms include transport, signaling, molecular motor, and gene regulation, among many others, and the knowledge of a protein's 3D structure is essential for understanding its function.
%The stable three-dimensional (3-D) structure of a protein is generally determined by its amino acid sequence. Predicting this 3D structure directly from the sequence, known as the protein folding problem, has been a central challenge across biology, chemistry, and biophysics for more than half a century.
Ever since Perutz's pioneering work, scientists have used laboratory techniques to determine the structures of ever-increasing numbers of proteins. Laboratory experiments for structure determination are labor-intensive, relying on methods such as X-ray crystallography, NMR spectroscopy, or, most recently, cryo-EM \citep[cryogenic electron microscopy,][]{yip2020atomic}. The resulting structures are often deposited in the publicly available Protein Data Bank \citep[PDB,][]{berman2000protein}, which has accumulated more than 245,000 entries to date. % throughout the last half-century and serves as essential training data for prediction methods.
Despite this wealth of structural data, high-throughput genome sequencing has, in comparison, generated hundreds of millions of protein sequences, of which fewer than 1\% have experimentally resolved structures \citep{bertoline2023before}.

Alongside historical developments in laboratory experiments, there was a growing scientific interest in what we now call the \textit{protein folding problem}: how does a protein, composed of a linear sequence of amino acids, acquire its stable 3D structure? On this question, Christian B.~Anfinsen, another Nobel laureate, posited that the stable 3D structure of a protein should be determined by its amino acid sequence \citep{anfinsen1973principles}. Thus, since laboratory structure determination could not keep pace with genome sequencing, the problem of computational protein structure prediction from its amino acid sequence gained widespread attention \citep{dill2012protein}. Over the years, computational methods have made incremental progress on this problem, as documented by the bi-annual  Critical Assessment of protein Structure Prediction (CASP, https://predictioncenter.org) experiments since 1994. During CASP experiments, participants submit their structure predictions under blinded conditions -- that is, the true structures of the target proteins are not disclosed at the time of submission. As a result, CASP provides a rigorous and valuable benchmark for assessing the predictive accuracy of different computational methods.
% However, a critical limitation remains: 
% Proteins are not always static. A growing class of proteins, known as metamorphic or fold-switching proteins, can adopt multiple distinct, stable 3D structures. Once thought to be rare [5, 6], these proteins are now believed to be more widespread [7, 8]. Their conformational changes, triggered by environmental cues or interactions with binding partners, are central to their function [9]. When applied to such proteins, AlphaFold2 typically predicts only one dominant conformation, missing alternative states [10].
% While AlphaFold3 [11] has improved the prediction of binding-induced conformational changes in complexes, a more profound challenge persists: identifying metamorphic proteins from their sequence alone and predicting their multiple apo (unbound) states. Solving this problem has significant implications. For instance, in drug design, a therapeutic targeting only one conformation of a metamorphic protein could be ineffective or even pathogenic if it stabilizes a disease-associated state [12].
A revolutionary breakthrough in structure prediction accuracy came with the arrival of AlphaFold2 \citep{jumper2021highly}, a transformer-based AI model developed by DeepMind to predict protein structure from a given amino acid sequence. AlphaFold2 achieved unprecedented accuracy of prediction at the atomic level in CASP14, and was subsequently recognized by the 2024 Nobel Prize in Chemistry.
This AI tool and its subsequent updates \citep{baek2021accurate, mirdita2022colabfold} have become a cornerstone of structural prediction, having now been used to predict hundreds of millions of structures \citep{varadi2024alphafold}.

The PDB, where each entry provides a protein sequence and its corresponding laboratory-determined structure, has served as essential ground-truth training data for AlphaFold2 and other structure prediction methods. However, many proteins -- particularly \textit{metamorphic} or \textit{fold-switching} proteins -- are not static/rigid entities. Instead, they are dynamic and capable of adopting multiple distinct 3D structures (or \textit{conformations}) in response to environmental factors, multimerization, and/or interactions with other molecules (e.g., binding partners) \citep{bu2011proteins}. 
Although once considered rare \citep{murzin2008metamorphic,bryan2010proteins}, an increasing number of metamorphic proteins have been discovered, indicating their population may be far more widespread than previously assumed \citep{lella2017metamorphic, porter2018extant}. Accurately identifying these metamorphic proteins and characterizing their distinct conformational states remains challenging for AI tools, including AlphaFold2, which was trained under the one-sequence-to-one-structure paradigm. As a result, AlphaFold2 has a critical limitation: when predicting the structures of known fold-switching or metamorphic proteins, which have at least two distinct yet stable conformations, by default, it can only predict one conformation, missing the other alternative structural states \citep{chakravarty2022alphafold2}.
AlphaFold3 \citep{abramson2024accurate} expanded AlphaFold2's functionality to the prediction of the structure of a protein together with its binding partners (i.e., the protein complex) and their binding-induced conformational changes. 

Important questions remain unanswered: (i) How can we  accurately determine whether a protein is metamorphic -- possessing at least two distinct conformations -- or single-fold, based solely on its amino acid sequence? (ii) How can we predict a protein's potential conformational changes independently of its binding partners? These questions represent a substantively difficult problem with profound applications \citep{chakravarty2025proteins}. 
For example, traditional drug design focuses on a single, static target protein. However, if the target protein can exist in multiple conformational states, a drug that targets one state may be ineffective or even harmful if the protein simply favors a pathogenic state as a result \citep{dishman2022design}. 
With fewer than 100 metamorphic proteins experimentally discovered to date \citep{porter2018extant}, computationally identifying all such proteins within the PDB and characterizing their potential conformational changes remains a formidable challenge.
Addressing this gap, which is the focus of this article, requires a comprehensive integration of powerful AI tools such as AlphaFold2 with effective statistical analysis.

\subsection{Basics of protein sequence and structure}
\label{subsec:sequence_to_structure}
\begin{figure}[h]
\centering
%\vspace{0.05cm}
\includegraphics[width=1\linewidth]{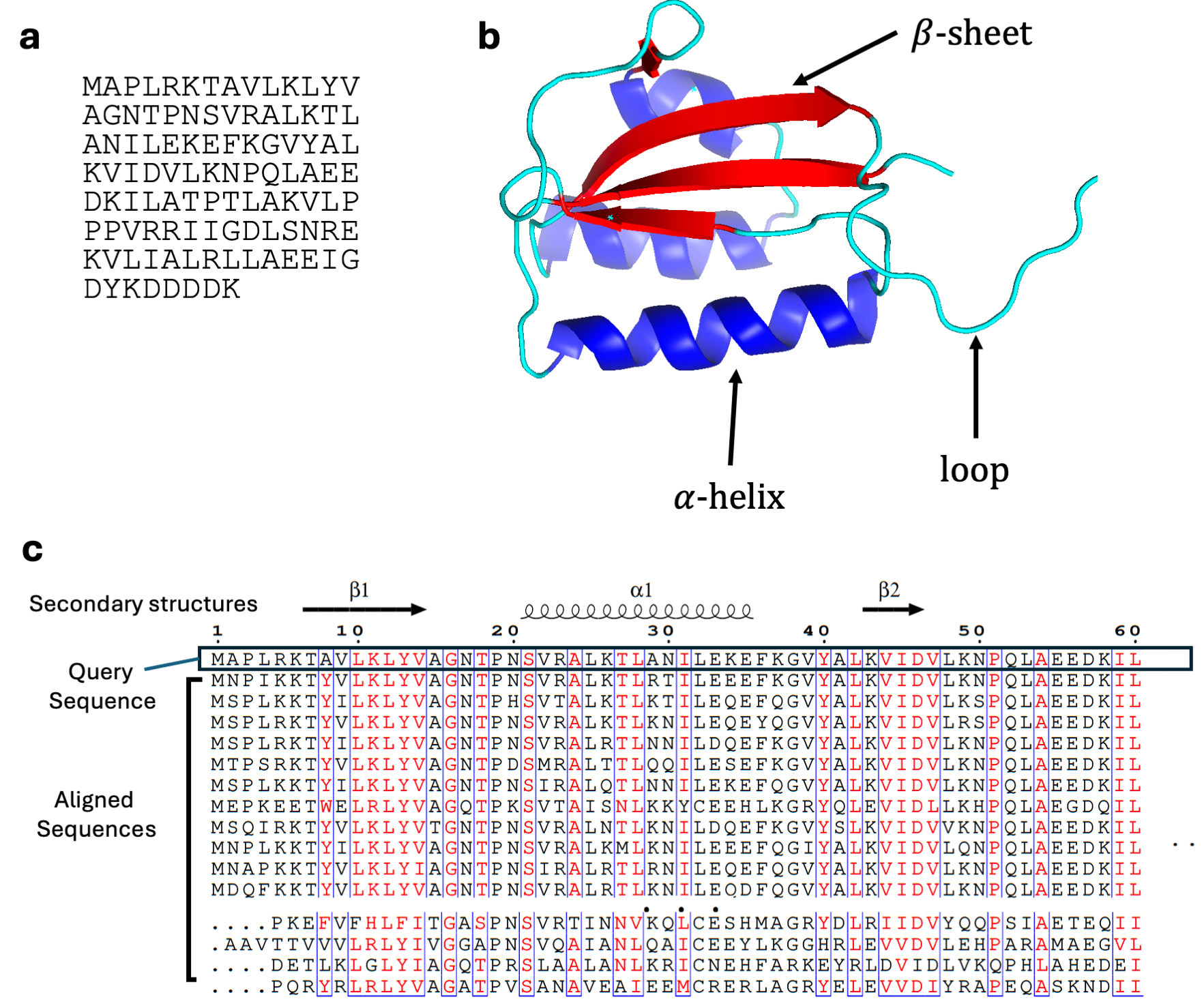}
\caption{(a). The amino acid sequence of the protein KaiB (106 residues). (b). 3D structure as determined by X-ray diffraction of Circadian Clock Protein KaiB in the native state (PDB ID: 2qkeE). The $\beta$-sheet segments are colored red, the $\alpha$-helix segments are colored blue, while the remaining segments that connect $\beta$-sheets and $\alpha$-helices are loops. (c). Multiple sequence alignments for KaiB using MMseqs2 \citep{steinegger2017mmseqs2} by querying the UniRef30 database \citep{suzek2015uniref}. A portion of sequences in the MSA is shown for amino acid positions 1-60. 
The secondary structures corresponding to each position are displayed above the sequences. The $\beta$-sheet is represented with an arrow, while the $\alpha$-helix is represented with a spring. Residues where sequence identity exceeds 0.9 are colored in red and framed in blue.
The MSA is visualized with ESPript3 \citep{gouet2003espript}.}
\label{fig:protein_bkg}
\end{figure}

A protein consists of a linear sequence of amino acids. As a concrete example, Fig.\ref{fig:protein_bkg} (a) displays the length 106 amino acid sequence of the Circadian Clock Protein KaiB \citep{chang2015protein}, where each letter represents one of the 20 different types of amino acids. An amino acid that is part of a protein sequence is also commonly referred to as a \textit{residue}; in general, the lengths of protein sequences can range from hundreds to thousands of residues.
A common way to represent protein structure, as in the PDB, is to specify the 3D Cartesian coordinates for the positions of each atom in the protein.
%The backbone of a protein consists of the interconnected sequence of N, $\text{C}_\alpha$, C, and O atoms corresponding to each amino acid,  
One 3D structure of KaiB is shown in Fig.\ref{fig:protein_bkg}(b), which was determined by X-ray diffraction and obtained from the PDB (ID: 2qkeE).
We highlight the two major secondary structure types, known as $\alpha$-helices and $\beta$-sheets, in the plotted 3D structure with blue and red colors, respectively. The segments of the protein that do not feature these regular secondary structures are known as loops or coils, as colored in light cyan.

\subsection{Proteins with Multiple Conformations}
\label{subsec:multiple_conformations}

% For decades, the challenge has been to learn the mapping from the space of amino acid sequences to the space of three-dimensional structures. The success of deep learning models like AlphaFold2 \citep{jumper2021highly} in approximating this mapping represents a monumental achievement in computational biology. However, this very success has refined the statistical question from ``What is the structure?'' to ``What is the \emph{distribution} of structures for a given sequence?'', acknowledging that the one-sequence-to-one-structure paradigm is an incomplete model of protein reality.

The foundational principle that guided protein structure prediction for many decades is that %of protein folding asserts that 
the true (or \textit{native}) conformation of a protein is uniquely encoded by its amino acid sequence \citep{anfinsen1973principles}. This principle assumes that a protein will stabilize at its lowest-energy conformation, in accordance with the energy landscape theory \citep{onuchic1997theory,wong2017fast,wong2018exploring}.
This classical view of protein folding implies that proteins have a single, static stable structure. This view has been challenged by the growing number of discoveries of metamorphic proteins -- proteins capable of folding into two or more distinctly different, stable conformations. 
One standing example of metamorphic proteins is the protein KaiB, whose conformational change will be presented in detail in Section \ref{sec:SMICE}.
From a statistical perspective, proteins that do not follow the traditional one-sequence-to-one-structure paradigm should instead be considered as being sampled from a multi-modal distribution within the conformational space. Consequently, the structure prediction problem should be reframed from ``What is the structure for a given sequence?'' to ``What is the distribution of possible structures?'' Identifying such metamorphic proteins and characterizing their conformational landscapes both remain significant challenges. Experimental methods are low-throughput and expensive, while current computational approaches have proven inefficient, suffering from low success rates due to insufficient data; see \cite{porter2024metamorphic} for a review.

\subsection{Multiple Sequence Alignment for Structure Prediction}
\label{subsec:msa_folding}
A key input that powers AlphaFold's protein structure prediction is %the evolutionary information encoded in
the multiple sequence alignment (MSA). 
An MSA is a collection of protein sequences that are believed to be evolutionarily or structurally related to the target sequence of interest.
These so-called homologous sequences are typically retrieved by querying large sequence databases based on hidden Markov models -- such as JackHMMer, HHblits, and mmseqs2 \citep{johnson2010hidden,remmert2012hhblits,steinegger2017mmseqs2} -- to find sequences similar to the query (target) sequence.
In Fig.\ref{fig:protein_bkg} (c), we present a portion of the MSA for KaiB, which contains 4,096 sequences, sorted by their identity (i.e., percentage of matching residues) to the query sequence.
For a query sequence of length $L$, an MSA, consisting of $N$ sequences, can be represented as $\mathcal{M} = \{\mathbf{Y}_1, \ldots, \mathbf{Y}_N\}$, where each $\mathbf{Y}_i$ is an $L \times 22$ binary matrix representing the one-hot encoding of the $i$-th homologous sequence. The 22 categories include the 20 standard amino acids, one for unknown amino acid types, and one for gaps in the alignment. 

An MSA can provide two types of evolutionary information that the AlphaFold model has likely learned to exploit for accurate structure prediction.
First, the conservation at each residue position in the sequence, revealed by the marginal distribution of amino acid types, provides an indication of the structural stability at that residue. 
If a sequence pattern over multiple residues appears repeatedly across many sequences, it typically corresponds to a specific structural motif; for example, the MSA of protein KaiB shows a highly conserved pattern around residues 10 to 15 (Fig.\ref{fig:protein_bkg} (c)), which corresponds to a $\beta$-sheet.

Second, the so-called coevolutionary information, revealed by statistical dependencies between residue pairs (sometimes far apart), often suggests that the highly dependent residue pairs are spatially close to each other in the folded structure. 
This may occur because a mutation at one residue often necessitates a compensatory mutation at its spatially-nearby partner residue to maintain the protein's structure and function.
With a well-constructed MSA as input, AlphaFold2 is generally still regarded to be the gold standard of protein structure prediction. Some newer protein language models that operate on the input sequence only (without any MSA), e.g., the ESM family \citep{hayes2025simulating} and OmegaFold \citep{wu2022high}, can have the advantage of being faster and simpler to use for structure prediction, but have not reached the same level of prediction accuracy as AlphaFold2.
% \subsection{AlphaFold and the Prediction of Protein Folding}
% \label{subsec:alphafold}
% A critical output of output is its confidence score (pLDDT), which provides a well-calibrated estimate of local prediction reliability. 

\subsection{Overview of Our Contribution and Method}\label{subsec:method}
In this article, our goal is to identify whether a given protein sequence is a metamorphic protein that can adopt multiple distinct folds, or a structurally stable, single-fold protein. To address this challenge, we developed a novel AI-driven classification approach that integrates the prediction power of AlphaFold2 with statistical learning. This integrated framework consists of three key steps: (1) conformational ensemble generation, (2) statistical feature extraction, and (3) binary classification.

Our pipeline begins with conformational ensemble generation, which is achieved through our previously developed MSA sampling method, SMICE (Sampling MSA Iteratively with CoEvolution information) \citep{chen2025uncovering}. As reviewed in Section \ref{sec:SMICE}, SMICE
repurposes AlphaFold2 from a single-structure predictor into a conformational ensemble predictor for a given target protein sequence. SMICE has demonstrated superior performance in generating diverse conformational ensembles for a benchmark set of metamorphic proteins, achieving high coverage of their distinct conformational states \citep{chen2025uncovering}.
Next, we perform statistical feature extraction on the predicted structure ensemble to characterize the modality of the conformational distribution. 
The statistical significance of these features for discrimination was rigorously validated on our carefully curated data sets of metamorphic and single-fold proteins.
Finally, we implement binary classification, using a random forest classifier to model the conditional probability of a protein being metamorphic given its vector of extracted features.

Our method achieved high predictive accuracy, with a mean area under the ROC curve (AUC) of 0.869 on the validation data set across 5-fold cross-validation, demonstrating its efficacy in discriminating between the two protein classes. 
To validate the practical application of our method, we applied the trained classifier to score over 600 proteins randomly selected from the Protein Data Bank. 
This application ranked proteins based on their predicted probability of being metamorphic, and identified several candidates with plausible conformational flexibility, thereby highlighting the method's potential for large-scale discovery.

By integrating AI‑driven protein structure prediction with statistical learning, our work offers an effective new strategy to identify metamorphic proteins and helps us understand how their structural flexibility contributes to molecular function. 
The rest of the article is organized as follows.
In Section \ref{sec:SMICE}, we present an overview of the conformational ensemble generation via SMICE with illustrative examples. The details of our method (and pipeline) are provided in Section \ref{sec:method}. In Section \ref{sec:results}, we present the classification results and the application of our classifier on 600 proteins sampled from the PDB database. We conclude the paper with a brief discussion in Section \ref{sec:conclusion}.
The technical details on implementation are provided in the Appendix.

\section{Generating conformational ensembles via MSA sampling and AlphaFold2}\label{sec:SMICE}
While AlphaFold2 was designed as a one-sequence-one-structure prediction tool, recent studies indicated that AlphaFold2 can be enhanced to provide predictions that capture multiple foldings for a protein sequence by modifying the input to AlphaFold2 to encourage its exploration of a broader range of the conformational landscape.
The most successful strategy to date is MSA sampling -- constructing a batch of smaller, shallower MSAs by selecting subsets of sequences from the full sequence alignment \citep{del2022sampling, monteiro2024high, wayment2024predicting, chen2025uncovering}, and then separately providing each subset as an input for AlphaFold2 to run.
When a structure is predicted for each of these distinct MSA subsets, the resulting conformational ensemble can potentially achieve a broader coverage of a protein's potential conformational states if high quality MSA inputs are supplied to AlphaFold2.

%From a statistical perspective
Recently, we proposed SMICE \citep{chen2025uncovering}, an approach that can be viewed as an iterative MSA sampling method from a statistical perspective. It formally embeds MSA sampling with generative probabilistic models and incorporates coevolutionary information (i.e., the statistical dependencies between the protein’s residue pairs, even when they are far apart) into the sampling criterion. 
Compared to other existing MSA sampling methods such as random sampling \citep{del2022sampling, monteiro2024high} and clustering \citep{wayment2024predicting}, SMICE's key advantages are its higher statistical efficiency and its utilization of coevolutionary information.
We found that the MSA subsets generated by SMICE yield conformational ensembles with high coverage of the conformational states on a benchmark set of metamorphic proteins. Moreover, SMICE incorporates a representative extraction procedure that not only clusters the predicted structures into groups based on their structure similarity but also identifies a representative structure for each cluster, enabling efficient characterization of the conformational landscape. We provide an overview of SMICE in the Supplementary Material Section \ref{supp:review_SMICE}. For a given protein sequence, SMICE outputs the tuple $(\{\mathcal{C}_k\}_{k=1}^{K}, \{S_k\}_{k=1}^{K})$, where $K$ is the total number of clusters identified, $\{\mathcal{C}_k\}_{k=1}^{K}$ denotes the set of structural clusters, and $S_k$ is the representative structure selected for cluster $k$. For each structure, in addition to the 3D arrangement of residues and atoms, SMICE also keeps and provides AlphaFold2’s confidence metric pLDDT (predicted local distance difference test), a built‑in measure by AlphaFold2 for its prediction confidence of the structure.

\begin{figure}[h]
\centering
%\vspace{0.05cm}
\includegraphics[width=1\linewidth]{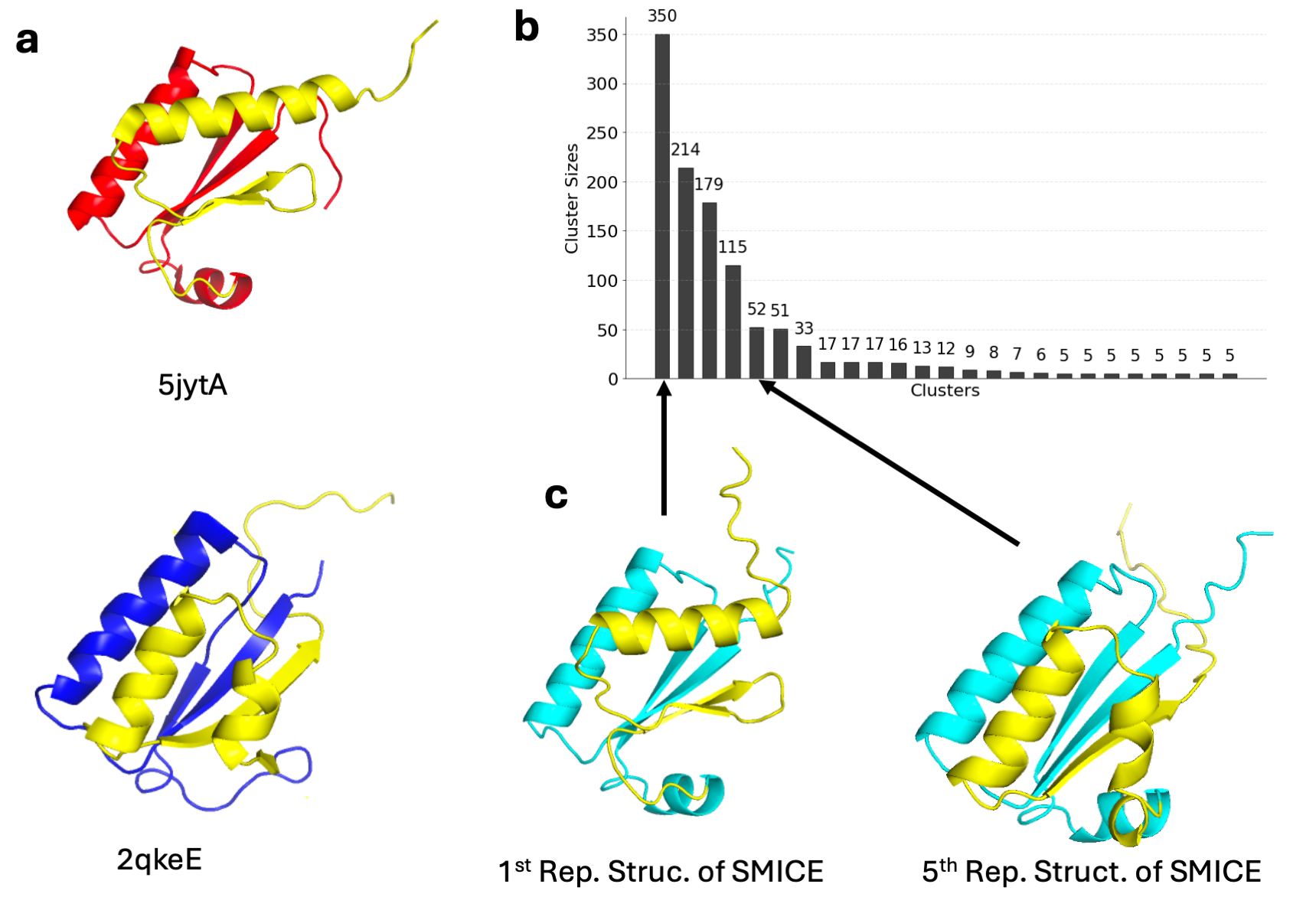}
\vspace{-0.5 cm}
\caption{(a). Crystal structures of KaiB in its two conformational states (PDB IDs: 5jytA and 2qkeE). Regions of conformational change identified by SMICE are colored yellow.
(b). The barplot of the cluster sizes for the conformational ensembles of protein KaiB (c). The representative structures that are closely matched to 5jytA (1st representative structure) and 2qkeE (5th representative structure).}
\label{fig:SMICE_exmp}
\end{figure}

In Fig.\ref{fig:SMICE_exmp}, we illustrate the application of SMICE to the metamorphic protein KaiB, which has two distinct experimentally confirmed conformational states (with PDB IDs 2qkeE and 5jytA, respectively). As shown in Fig.\ref{fig:SMICE_exmp}(a), there is a significant change in the secondary structure (localized in the C-terminal domain) between these two states.
Fig.\ref{fig:SMICE_exmp} (b) displays the sizes of the clusters of the conformational ensembles predicted by SMICE. 
SMICE identified two representative structures that well capture the conformations of KaiB: the first representative structure (cluster size 350) closely matches structure 5jytA, and the fifth representative structure (cluster size 52), closely matches structure 2qkeE as shown in Fig.\ref{fig:SMICE_exmp} (c). In contrast, AlphaFold2 by default can only find one structure: 5jytA.
%\textcolor{red}{\bf [[Please check the previous sentence.]]}

% To analyze the structural ensemble generated by SMICE, we first computed the contact map for each predicted structure, which is the distance matrix between all $C_\alpha$ atoms, and applied PCA transformation on the vectorized contact maps. 

% In Fig.\ref{fig:SMICE_exmp} (a), we visualize the PCA embedding of the predicted conformational ensembles, where each point represents a single predicted structure, colored by the TMscore difference (TMscore1-TMscore2).
% We found that the predicted conformational ensembles cleanly separate into two major classes, with each class corresponding closely to one of the two KaiB conformations. 
% What's more, SMICE automatically identifies the region of highest conformational variability.

% This is achieved by computing a variance matrix where each entry corresponds to the variance of the contact distance between corresponding residues across the ensemble. The region of which corresponding submatrix has the highest variance is identified as the variable region, as shown in Fig.\ref{fig:SMICE_exmp} (b).

\section{Classification of metamorphic and single-fold proteins}\label{sec:method}
\subsection{Training data construction}
Given the scarcity of experimentally validated metamorphic proteins -- fewer than 100 have been identified to date -- and the likelihood that many proteins might possess undiscovered alternative folds, constructing a reliable training dataset is important, particularly in avoiding the mislabeling of metamorphic proteins as single‑fold.

%, while selecting single-fold proteins are selected from a large population to ensure their representative.

We use a benchmark set of 92 experimentally verified fold-switching proteins provided in \citet{chakravarty2024alphafold} as the dataset for metamorphic proteins. 
To construct the dataset for the class of single-fold proteins, we curated structurally stable proteins from two databases: ATLAS \citep{vander2024atlas} and CoDNaS-Q \citep{escobedo2022codnas}. 

The ATLAS database systematically collects all-atom molecular dynamics (MD) simulations for a large set of proteins. MD simulation is a computational technique that simulates the physical movements of atoms over time \citep{karplus1990molecular, hollingsworth2018molecular}. By numerically solving the equations of motion (under the force field), it generates a detailed trajectory that depicts how a protein's structure evolves from a starting conformation according to the energy landscape \citep{onuchic1997theory}.
To quantify local structural flexibility from the simulated trajectory, the root mean square fluctuation (RMSF) for a given residue is calculated as the square root of the time-averaged squared distance between the residue's positions and its time-averaged position.
A protein's global structural stability is assessed by averaging the RMSF values across all its residue positions, and a protein with a low average RMSF suggests high energetic stability and a single, dominant conformational state. 
Based on this criterion, we selected the top 200 proteins with the lowest average RMSF values from the entire ATLAS database to include in our set of structurally stable, single-fold proteins. %\textcolor{red}{\bf [[Please give an explanation or citation of why 0.83 \r{A} is chosen]]}

CoDNaS-Q (Conformational Diversity of Native State – Quaternary) provides a complementary, experimentally derived measure of structural stability. It groups protein structures from the PDB that share high sequence identity, treating them as different experimental observations of the same protein. To quantify a protein's structural variability across these observations, the root-mean-square deviation (RMSD), calculated as the square root of the averaged squared distance between the corresponding atoms of two aligned structures, is computed for every pair of these experimentally observed structures. A protein's maximum pairwise RMSD reflects the largest observed structural deviation in its experimental record. Based on this criterion, we selected the top 200 proteins with the lowest maximum RMSD values to include in our set of experimentally stable, single-fold proteins. %\textcolor{red}{\bf [[Please give an explanation or citation of why 178 is chosen]]}

Combining the single‑fold proteins identified from the ATLAS and CoDNaS‑Q databases allows for a more comprehensive assessment of structural stability when defining single‑fold proteins. ATLAS selects proteins that are intrinsically stable \textit{in silico}, based on the principles of molecular physics and energy landscapes. The CoDNaS-Q database identifies proteins that are consistent \textit{in vitro}, maintaining a stable fold across diverse experimental conditions. 

For a given protein, if its full MSA set contains only a small number of sequences, there is typically insufficient information to evaluate its structural variability. With this in mind, we removed proteins whose full MSAs contained fewer than 20 sequences. After this filtering step, we obtained in our training dataset 80 metamorphic proteins (each exhibiting two distinct conformations with a pairwise RMSD greater than 4 Å), 128 single‑fold proteins from ATLAS with an average RMSF below 0.83 \r{A}, and 178 single‑fold proteins from CoDNaS‑Q with a maximum RMSD below 0.80 \r{A}. These single‑fold proteins indeed show minimal structural variation, especially considering that the van der Waals radius of an atom is 1–2 \r{A} (e.g., 1.7 \r{A} for carbon) \citep{bondi1964van}.

\subsection{Extracting features from predicted ensembles}
By running SMICE on the 80 metamorphic proteins and 306 single-fold proteins selected from ATLAS and CoDNaS-Q, we obtained the predicted conformational ensembles and the representative structures for each protein. For the $i$th protein, we denote the SMICE output as the tuple $( \{\cC_k\}_{k=1}^{K_i},  \{S_k\}_{k=1}^{K_i})$, where $K_i$ is the number of clusters, $\{\cC_k\}_{k=1}^{K_i}$ represents the set of structural clusters, ordered by decreasing size such that $\left|\cC_1\right| \geq\left|\cC_2\right| \geq \cdots \geq\left|\cC_K\right|$,  and $\{S_k\}_{k=1}^{K_i}$ denotes the corresponding representative structures selected for each cluster.

% We performed Mann-Whitney U tests on key features and provided the Benjamini-Hochberg corrected p-values with the comparative boxplots of metamorphic versus single-fold proteins in Fig.\ref{fig:features}.

\begin{figure}[tb]
\centering
%\vspace{0.05cm}
\includegraphics[width=1\linewidth]{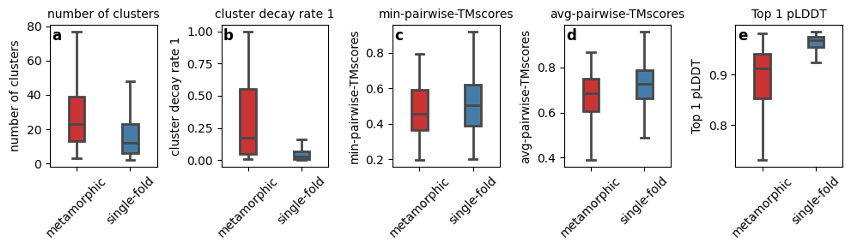}
\vspace{-0.5 cm}
\caption{Feature comparison between metamorphic and single-fold proteins derived from SMICE conformational ensemble predictions.}
% Each subplot displays the distribution of one feature with its corresponding FDR-adjusted p-value from Mann-Whitney U tests. (h).
% The scatterplot of the number of clusters against the sequence length. The Spearman correlation coefficient and associated p-value are provided.}
\label{fig:features}
\end{figure}

A key feature to consider is the modality of the conformational ensemble, when viewed as a probability distribution over states. 
The predicted conformational ensemble of a single-fold protein is expected to concentrate around one dominant conformation, with rapidly decaying cluster sizes. 
In contrast, metamorphic proteins should exhibit multiple well-populated clusters distinguished by substantial structural dissimilarity.
These expectations are supported by contrasting the number of clusters and the cluster decay rate $R_1 = |\mathcal{C}_2|/|\mathcal{C}_1|$ between metamorphic and single-fold proteins, as shown in Fig.\ref{fig:features} (a)-(b). Consequently, we include both the number of clusters and cluster size decay rate $R_1$ as predictive features. To provide more comprehensive information, we also include additional decay rates $R_2$ and $R_3$, which capture the size‑decay patterns of subsequent clusters, as predictive features. Here $R_{k} = |\mathcal{C}_{k+1}|/|\mathcal{C}_k|$.

Additionally, we use pairwise structural dissimilarity to directly quantify the diversity of the selected representative structures $\{S_k\}_{k=1}^{K_i}$. 
While single-fold proteins tend to exhibit high structural similarity across all pairs of representative structures (i.e., the structural clusters identified are not really that different), metamorphic proteins tend to have low structural similarity across clusters. 
To assess similarity between each pair of structures, we adopt the template modeling score (TMscore) \citep{zhang2004scoring}, a widely used metric that measures the similarity between two structures with a score between 0 and 1, with 1 indicating a perfect match and lower value indicating greater dissimilarity. 
The TMscore provides an interpretable measure of similarity across different proteins.
To quantify structural divergence between clusters, we computed the minimum and average pairwise TMscores among the representative structures $\{S_k\}_{k=1}^{K_i}$ in the SMICE-predicted conformational ensemble. 
The minimum TMscore corresponds to the two most structurally distinct representatives generated by SMICE.
As shown in Fig.\ref{fig:features} (c)-(d), metamorphic proteins show significantly lower values for both minimum and average pairwise TMscores compared to single-fold proteins. This difference supports the intuition that the SMICE-predicted ensembles of metamorphic proteins exhibit greater structural dissimilarity compared to single-fold proteins.

Finally, AlphaFold2's confidence metrics -- AlphaFold2’s built-in pLDDT (predicted local distance difference test) score -- are also used as predictive features. We found that AlphaFold2 tends to be less confident about its predictions on protein sequences with multiple conformations.
As shown in Fig.\ref{fig:features}(e), the highest pLDDT values among representative structures $\{S_k\}_{k=1}^{K_i}$ are substantially lower for metamorphic proteins compared to single-fold proteins. This phenomenon may be explained as, although AlphaFold2 was trained under a one‑sequence‑one‑structure paradigm, the existence of multiple conformations in metamorphic proteins reduces its confidence in generating a single structural prediction.
To be sufficiently informative and robust, we included the three highest pLDDT values from the representative structures as features (for proteins with fewer than three representative structures, the unavailable pLDDT entries were coded as NA, i.e., missing).

% Finally, the side information of the protein sequence and MSA serve as controlling factors.
% MSA depth directly affects the reliability and diversity of the evolutionary information leveraged by SMICE's MSA sampling step. 
% Furthermore, we observed that metamorphic proteins tend to have shallower MSAs, as shown in Fig.\ref{fig:features} (f). 
% We also consider the sequence length as an important controlling factor.
% Although sequence length itself is not a powerful predictor of metamorphicity as shown in Fig.\ref{fig:features} (g), it could be a significant confounder since it is correlated with the number of clusters, which is a key predictor (Fig.\ref{fig:features} (h)).

Table \ref{tab:feature_summary} presents a summary of the features we extracted from SMICE-predicted conformational ensembles.

\begin{table}[tbp]
    \centering
    \caption{Summary of features extracted from the SMICE conformational ensemble for classifying single-fold versus metamorphic proteins.}
    \label{tab:feature_summary}
    \begin{tabular}{p{3cm}p{9cm}}
        \toprule
        \textbf{Feature Category} & \textbf{Features and Descriptions} \\
        \midrule
        Representative extraction results & 
        \makecell{• Number of structural clusters ($K$) \\
        • Decay rates of the cluster sizes: $\{R_k=|\mathcal{C}_{k+1}|/|\mathcal{C}_{k}|\}_{k=1}^3$} \\
        \addlinespace
        Pairwise structural dissimilarity & 
        \makecell{• Minimum pairwise TMscore among all representative structures \\
        • Average pairwise TMscore among all representative structures} \\
        \addlinespace
        AlphaFold2's confidence metric & 
        • pLDDT values of the top 3 highest-confidence representative structures \\
        \bottomrule
    \end{tabular}
\end{table}
% \subsection{Multi-conformers identification} 
\subsection{Random Forest Model for Classification}
\label{subsec:classifier}
Let $\mathcal{D} = \{(\mathbf{x}_i, y_i)\}_{i=1}^n$ denote the training dataset, where $\mathbf{x}_i \in \mathbb{R}^{9}$ is the vector of extracted features for the $i$th protein's predicted ensembles from SMICE (Table~\ref{tab:feature_summary}) and the response $y_i \in \{0, 1\}$ corresponds to a metamorphic protein ($y_i=1$) or a single-fold protein ($y_i=0$). 
We consider the random forest model for modeling the conditional probability,
\begin{equation}\label{eq:RF}
    P(y_i=1|\mathbf{x}=\mathbf{x}_i) \equiv \text{RF}(\bx_i; \{\bstheta_b\}_{b=1}^B) = \frac{1}{B} \sum_{b=1}^B f(\mathbf{x}_i;\bstheta_b),  
\end{equation}
where $f{(\mathbf{x};\bstheta_b)} \in [0,1]$ represents the classification probability predicted by the $b$th classification tree, and $\bstheta_b$ represents its parameter, including the tree structure, the decision rules of interior nodes, and the parameter associated with the terminal nodes. 

To mitigate bias toward the majority class of single-fold proteins, we estimate the parameters $\{\bstheta_b\}_{b=1}^B$ by maximizing the balanced accuracy, i.e., the average of the true positive rate (TPR) and true negative rate (TNR),
%For a given decision threshold $\tau$, the balanced accuracy (BA) is defined as the arithmetic mean of sensitivity and specificity:

\begin{equation}\label{eq:balanced_accuracy}
\begin{aligned}
\frac{1}{2} \left(
\frac{\sum_{i=1}^n y_i \mathbb{I}(\text{RF}(\bx_i; \{\bstheta_b\}_{b=1}^B) > \tau)}{\sum_{i=1}^n \mathbb{I}(y_i = 1)} +
\frac{\sum_{i=1}^n (1-y_i) \mathbb{I}(\text{RF}(\bx_i; \{\bstheta_b\}_{b=1}^B) \leq \tau)}{\sum_{i=1}^n \mathbb{I}(y_i = 0)}
\right),
\end{aligned}
\end{equation}
where $\mathbb{I}(\cdot)$ is the indicator function.  The classification threshold $\tau$ is tuned to balance the TPR and TNR.
Moreover, the hyperparameters of the random forest model, such as the number of trees $B$, the maximum depth of each tree, were selected via a 5-fold cross-validation procedure combined with an exhaustive grid search. 
 The optimal hyperparameter with the highest average balanced accuracy across the five validation folds was selected. 
 Finally, a random forest model was refit using this optimal hyperparameter configuration on the entire training dataset.

Missing (NA) values in the top-3 pLDDT features were handled in our implementation of the random forest classification by using the Missing Incorporated in Attributes (MIA) approach, where a missing value is treated as an extra category for node splitting in the trees \citep{twala2008good}.

\section{Application and Results}\label{sec:results}
\subsection{Classifying metamorphic and single-fold proteins}
To evaluate the performance of the random forest classifier, we employed 5-fold stratified cross-validation by randomly partitioning the full dataset into training and validation sets while preserving class proportions. For each fold, a random forest model was trained following the procedure discussed in Section \ref{subsec:classifier} and evaluated on the validation set. 

\begin{figure}[tbp]
\centering
%\vspace{0.05cm}
\includegraphics[width=1\linewidth]{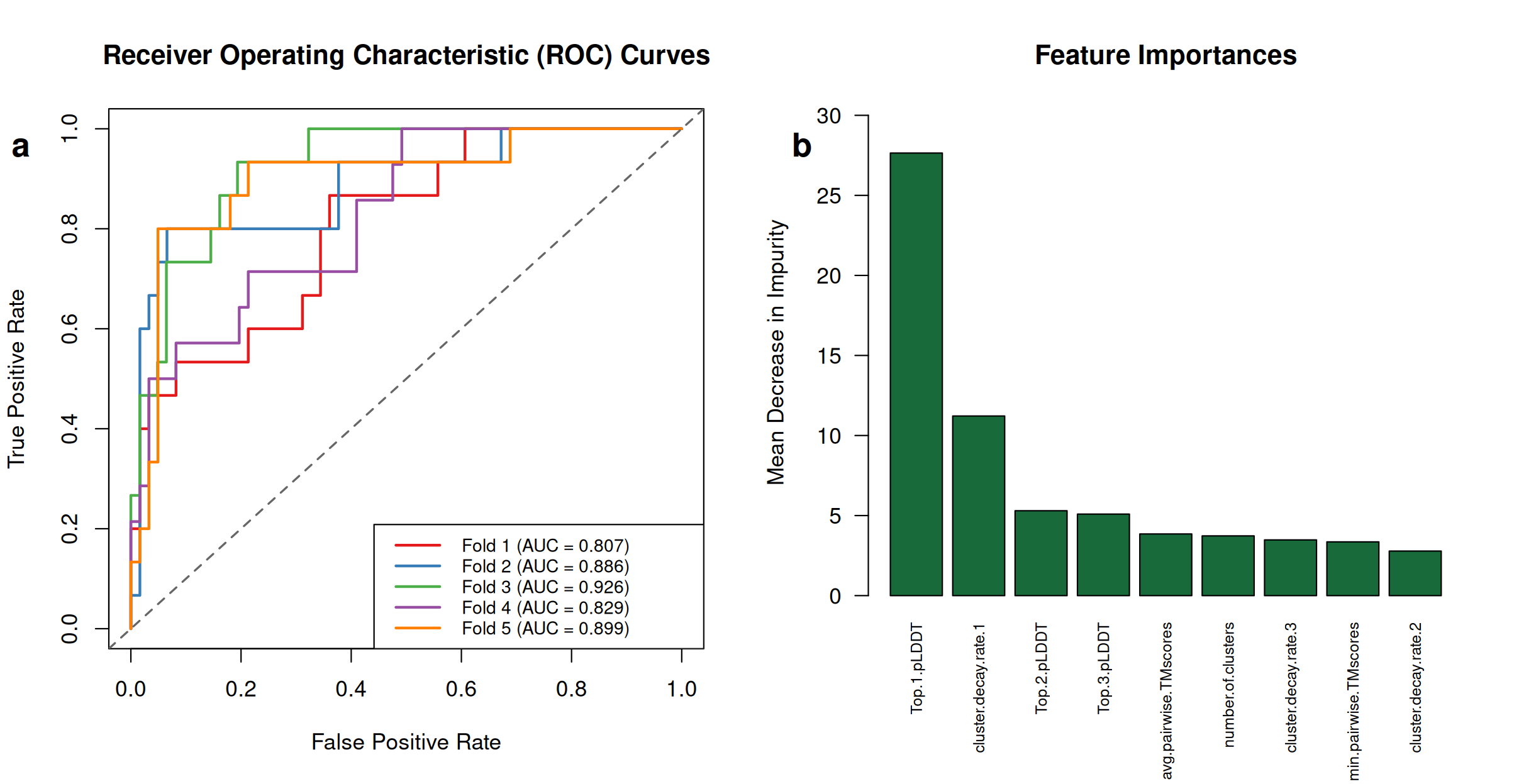}
\vspace{-0.5 cm}
\caption{(a) ROC curves on validation dataset from 5-CV folds with corresponding AUC values. (b) Feature importance rankings based on the mean decrease in impurity of the random forest.
}
\label{fig:classification}
\end{figure}

The random forest classifier demonstrated good performance in classifying metamorphic proteins from single-fold proteins by achieving a mean AUC of 0.869 (standard deviation = 0.050) across 5-fold CV, as shown in Fig.\ref{fig:classification} (a).
In particular, the model achieved high specificity for single-fold proteins (mean TNR = 0.775 with standard deviation 0.058 across 5-fold CV) and high sensitivity (mean TPR = 0.796 with standard deviation 0.185) for metamorphic proteins, under the selected classification threshold of $\tau = 0.18$.

We subsequently trained the random forest classifier on the full dataset as our working model. A ranking of features' importance, based on the mean decreases in impurity of this working model, is shown in Fig.\ref{fig:classification} (b). 
The top pLDDT value (i.e., the highest pLDDT among the representative structures) emerged as the most predictive feature, suggesting that AlphaFold2's built-in confidence score meaningfully captures some of the ambiguity %inherent in training 
associated with sequences %having 
that could have multiple conformations. 
The next most important features included the cluster decay rates, the number of clusters, and the averaged and minimum pairwise TMscores, highlighting that the existence of multiple well-populated clusters as uncovered by SMICE is a key indicator of metamorphic proteins.

\subsection{Discovering metamorphic proteins from the PDB database}
To assess the potential of our approach for discovering metamorphic proteins, we randomly selected 600 proteins from the PDB database. 
We ran SMICE on each protein to generate conformational ensembles, extract features, and compute the predicted probability of being a metamorphic protein using the working model of our random forest classifier. 
These 600 proteins were ranked based on their predicted probabilities of being metamorphic.

Table \ref{tab:top5_metamorphic_predictions} lists the top five proteins that our method predicts to have the highest probability of being metamorphic. 
A common feature among these proteins is their binding roles in dynamic molecular interactions. For example,  protein 3j07R (Alpha-crystallin B chain) is a chaperone that binds diverse client proteins \citep{jehle2011n}; 5gmkG (CWC25) is a splicing factor that interacts with mRNA and the spliceosome \citep{chiu2009cwc25}.
This strong association with ligand or partner binding supports our prediction, as conformational changes are often required for proteins to transit between bound and unbound states or to accommodate different binding partners.

\begin{table}[tbp]
\centering
\caption{Top five proteins predicted to be metamorphic with highest probability.}
\label{tab:top5_metamorphic_predictions}
\begin{tabular}{p{0.9cm}p{2.1cm}p{2cm}p{3cm}p{1.7cm}}
\toprule
\textbf{PDB ID}  & \textbf{Organism} & \textbf{Protein} & \textbf{Function} & \textbf{Predicted Probability}  \\
\midrule
4d5le & Oryctolagus cuniculus & 40S ribosomal protein S30 & structural constituent
of ribosome; antibacterial humoral response & 0.927\\
\addlinespace
4v4gF1& Escherichia coli  & 50S ribosomal protein L31 & structural constituent of ribosome; zinc ion binding; rRNA binding& 0.904\\
\addlinespace
3j07R& Homo sapiens & Alpha-crystallin B chain & 	chaperone; amyloid-beta binding; 	identical protein binding; metal ion binding; 	microtubule binding; etc.  & 0.900 \\
\addlinespace
5gmkG& Saccharomyces cerevisiae S288C  & Pre-mRNA-splicing factor CWC25 & mRNA processing; mRNA splicing &  0.899\\
\addlinespace
4wu1I5&  Thermus thermophilus HB8, Escherichia coli &  50S ribosomal protein L31 &  structural constituent of ribosome; zinc ion binding; rRNA binding & 0.897 \\
\bottomrule
\end{tabular}
\end{table}

\begin{figure}[tb]
\centering
%\vspace{0.05cm}
\includegraphics[width=1\linewidth]{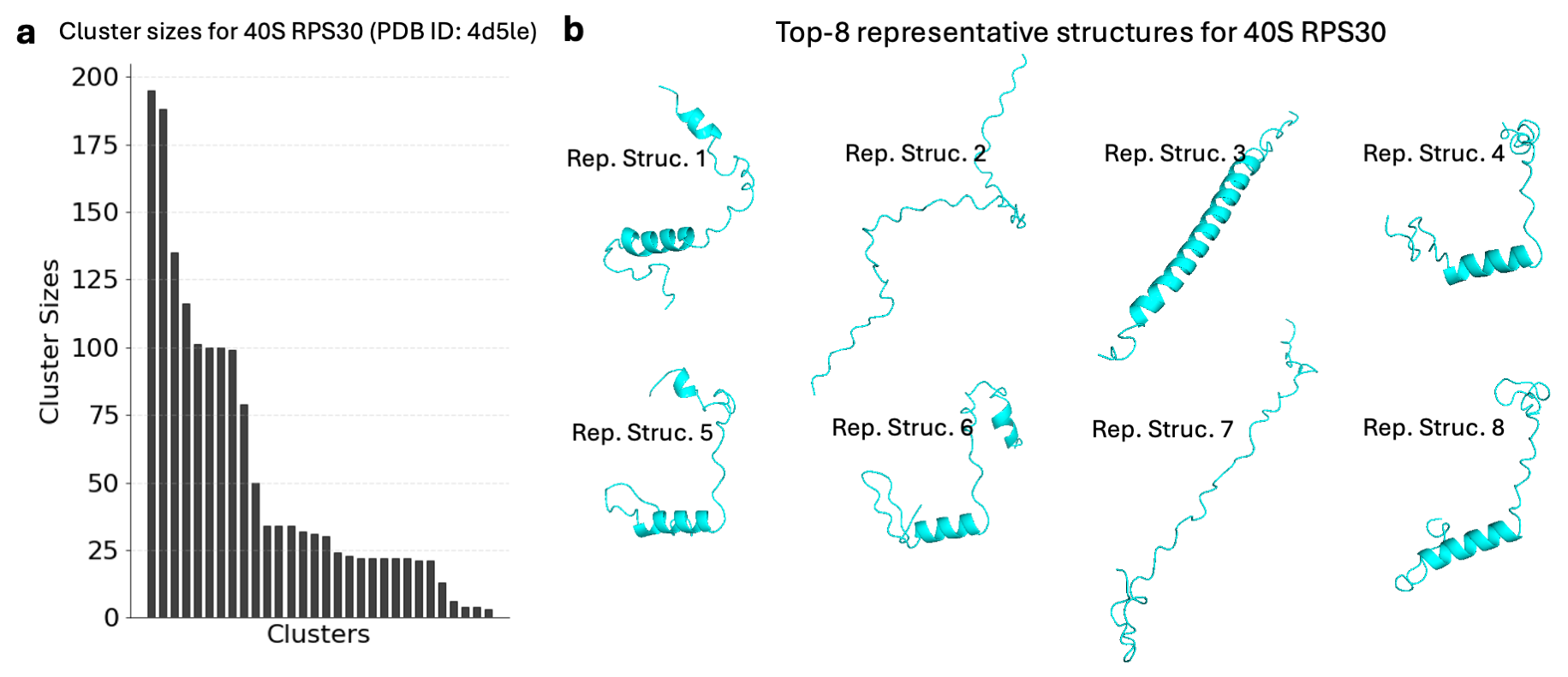}
\vspace{-0.5 cm}
\caption{(a) The cluster sizes of the predicted conformational ensemble for 40S ribosomal protein S30.
(b) The representative structures of the eight largest clusters extracted by SMICE for 40S ribosomal protein S30. 
}
\label{fig:exmp_4d5le}
\end{figure}

The top-ranked metamorphic protein candidate predicted by our method is the 40S ribosomal protein S30 (RPS30; PDB ID: 4d5le), a 59-amino acid subunit essential for mRNA translation. Our method predicted that this protein has a 0.927 probability of being metamorphic. 
 SMICE extracted 29 clusters from the predicted conformational ensemble for RPS30, with the cluster sizes showing slow decay as depicted in Fig.\ref{fig:exmp_4d5le} (a). By visualizing the representative structures of the eight largest clusters, we found considerable dissimilarity among the predicted conformations, including purely random coil conformations, such as representative structures 2 and 7, and more ordered conformations containing varying amounts of $\alpha$-helical segments. 
This finding is consistent with the understanding of the dual functionality of RPS30.
In addition to its well-characterized role as a structural constituent of the ribosome \citep{rabl2011crystal}, some studies reveal the bactericidal effects of RPS30 as an antimicrobial peptide in combating drug-resistant bacteria and mediating the host's innate immune response \citep{tollin2003antimicrobial,brouwer2006synthetic}.
A recent study found that this function is attributed to RPS30's preferential binding to bacterial membranes \citep{bhatt2025unravelling}.
Upon binding bacterial membranes, it undergoes a conformational transition from a random coil to an $\alpha$-helix. In contrast, interaction with mammalian membranes does not induce this helical conformation.
%where RPS30 undergoes a conformational transition from a random coil to an $\alpha$-helix. In contrast, RPS3 remains as a random coil in the mammalian membrane\citep{bhatt2025unravelling}.

\begin{figure}[h]
\centering
%\vspace{0.05cm}
\includegraphics[width=1\linewidth]{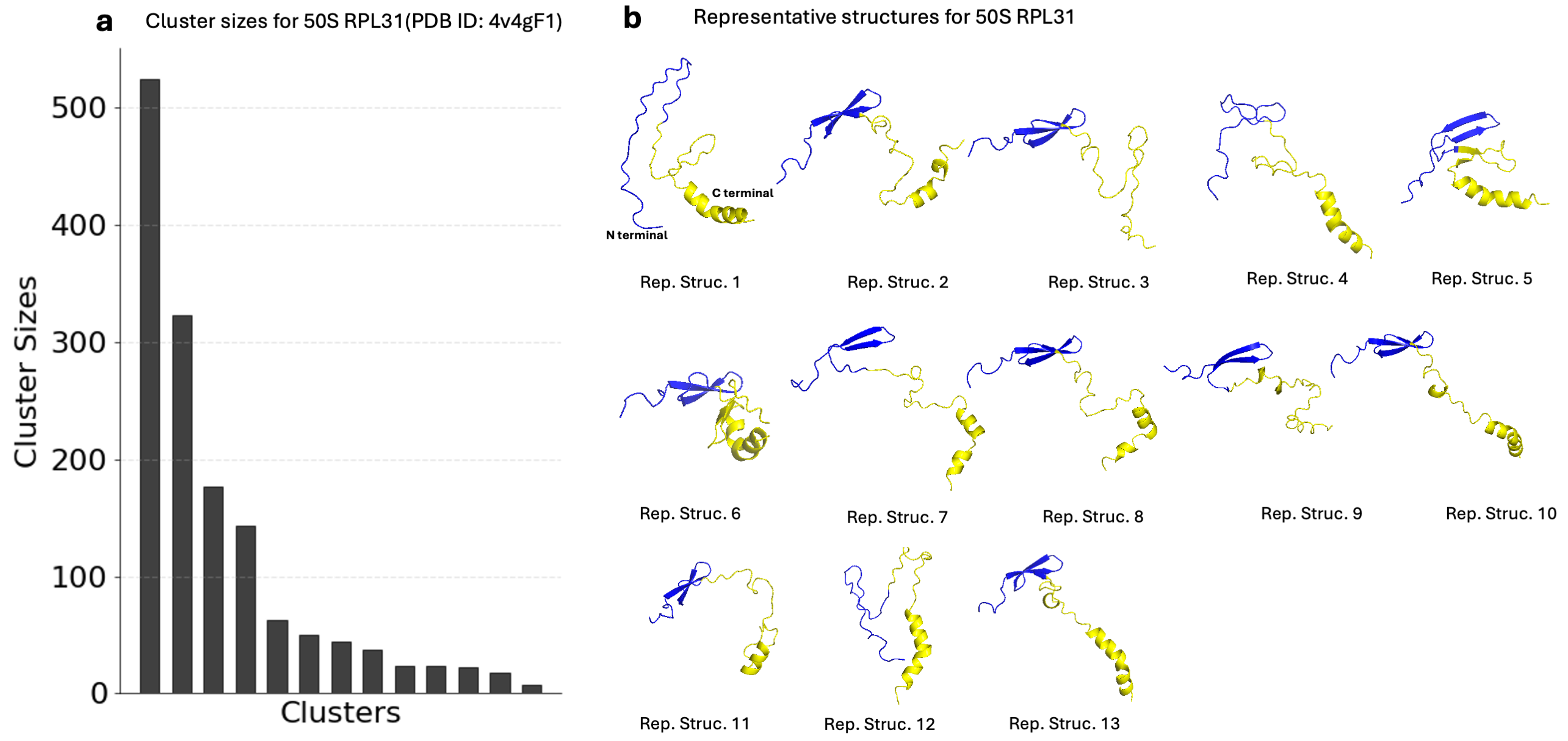}
\vspace{-0.5 cm}
\caption{(a) The cluster sizes of the predicted conformational ensemble for 50S ribosomal protein L31. (b) The representative structures of the thirteen clusters extracted by SMICE for 50S ribosomal protein L31. The identified regions of highest variability are colored in yellow.
}
\label{fig:exmp_4v4gF1}
\end{figure}

The second top-ranked metamorphic protein candidate by our method is the 50S ribosomal protein L31 (RPL31, PDBID: 4v4gF1). Our method predicted that this protein has a 0.904 probability of being metamorphic. In \textit{Escherichia coli}, RPL31 acts as a flexible bridge connecting a large 50S ribosomal protein subunit and a small 30S ribosomal protein subunit, forming the 70S ribosome.
As shown in Fig.\ref{fig:exmp_4v4gF1} (a), SMICE identified 13 clusters in the predicted conformational ensemble of RPL31, with representative structures displayed in Fig.\ref{fig:exmp_4v4gF1} (b).
The region of highest variability, as identified by SMICE, consists of structural elements that shift between a loop (representative structures 1,2,4,7,8,11), an $\alpha$-helix (representative structures 9, 10, 13), or a $\beta$-sheet segment (representative structures 5, 6), adjacent to a C-terminal $\alpha$-helical tail. 
The identified variable region corresponds to the linker region of RPL31.
By analyzing atomic models for the RPL31 of the 70S ribosome from Escherichia coli, \cite{fischer2015structure} found the linker region of RPL31 had a distinct conformational change as the 30S ribosome ratchets during translation elongation.
The switch occurred between an extended conformation,  where loops connect the N-terminal $\beta$-sheet head to the C-terminal $\alpha$-helical tail, and a kinked conformation, where an $\alpha$-helix formed the connection.
Furthermore, RPL31 has extraribosomal functions, including autoregulation via RNA binding \citep{bressin2019tripepsvm} and serving as a Zn$^{2+}$ reservoir in the cell \citep{hensley2012characterization}. Both RNA and zinc binding are likely to induce additional conformational changes in the protein.

The identification of metamorphic proteins by our method is thus supported by current biological understanding of their roles and structures.

\section{Conclusion}\label{sec:conclusion}
In this paper, we presented an approach to address the long-standing challenge of identifying metamorphic (fold-switching) proteins, a critical problem in structural biology that challenges the traditional one-sequence-one-structure view of protein folding.

By leveraging the predictive power of AI-based AlphaFold2 with a suite of statistical methodologies, we developed a new classification framework for determining if a given protein is metamorphic or single-fold based solely on its amino acid sequence.
The classifier utilizes our developed MSA sampling method, SMICE, which repurposes AlphaFold2 from a single-structure predictor into a conformational ensemble predictor for exploring the conformational landscape.
From the resulting ensembles, we systematically characterized and extracted statistical features, including the ensemble's modality, structural diversity, and model confidence of the predicted conformational ensemble. 
With a carefully curated dataset comprising known metamorphic proteins and single-fold proteins, we identify the extracted features having high statistical significance for the classification task. 
A random forest classifier trained on these features achieved high accuracy (AUC = 0.869) for differentiating between metamorphic and single-fold proteins.
The application of this classifier to the Protein Data Bank identified several candidate proteins with plausible conformational flexibility, highlighting the method's potential for accelerating the discovery of novel metamorphic proteins, which can help develop new biosensors or targeted drug delivery systems.

%Further study is ongoing in developing uncertainty quantification tools for the classifier's predictions.
%What's more, 
One future direction is to improve the scalability of the classification pipeline through more efficient sampling methods, which is a necessary step for a comprehensive exploration of the entire Protein Data Bank, which hosts hundreds of thousands of proteins. 

Our work demonstrates that effective statistical analysis can substantially enhance AI-driven tools in applied science by
providing unique insights and sound statistical reasoning.
We anticipate more success in integrating statistical methods with modern AI development in scientific and engineering applications.

\section*{Data and Code availability}
The code corresponding to SMICE is publicly available at GitHub (\url{https://github.com/StatCYK/SMICE}).
The curated datasets and the code corresponding to the classifier are publicly available at GitHub (\url{https://github.com/StatCYK/Metamorphic-Classify}).

\section*{Acknowledgements}
S.W.K.~Wong's research is supported in part by Discovery Grant RGPIN-2019-04771 from the Natural Sciences and Engineering Research Council of Canada. S.~C.~Kou acknowledges support from Harvard Data Science Initiative (HDSI).
\bibliography{ref}

\newpage
\appendix

\setcounter{page}{1} 
\setcounter{equation}{0}
\setcounter{figure}{0}
\setcounter{table}{0} 
\renewcommand{\thetable}{S.\arabic{table}}
\renewcommand{\thefigure}{S.\arabic{figure}}
\renewcommand{\thesection}{S.\arabic{section}}
\renewcommand{\theequation}{\thesection.\arabic{equation}}
\renewcommand{\thealgorithm}{S.\arabic{algorithm}}

 \begin{center}
   {\Large\bf Supplementary Material for  ``Classifying Metamorphic versus Single-Fold Proteins with Statistical Learning and AlphaFold2''}  
 \end{center}

\section{Review of SMICE}\label{supp:review_SMICE}
In this section, we present an overview of SMICE, the method we developed for predicting the multiple conformations of metamorphic proteins. See \cite{chen2025uncovering} for the full details of the implementation.

\begin{figure*}[bp]
%\vskip 0.2in
\centering
\includegraphics[width=0.85\linewidth]{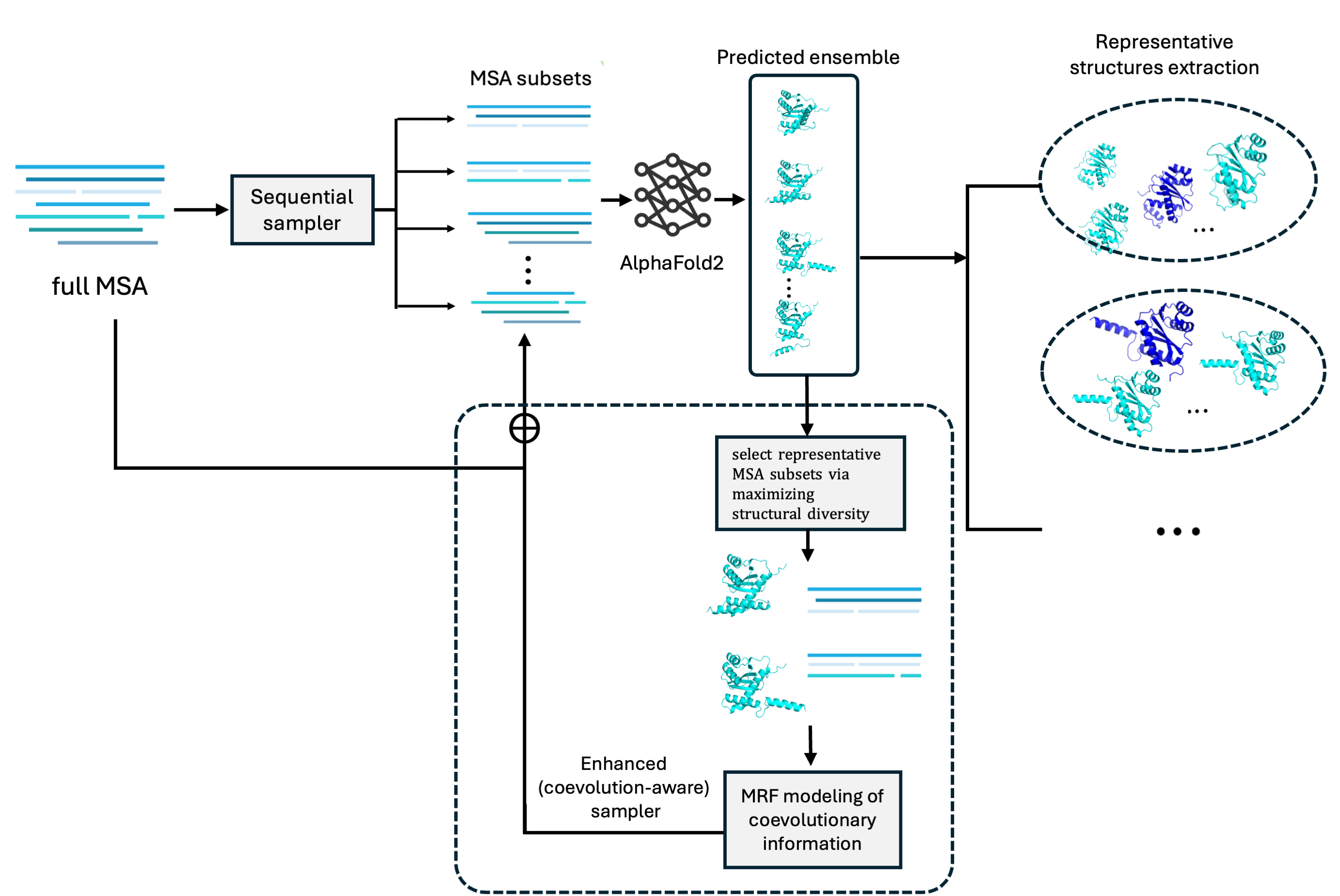}
\vspace{-0.2cm}
\caption{SMICE workflow. In SMICE's sampling step, MSA subsets are drawn from the full MSA using sequential sampling. Then, structure predictions are made on the MSA subsets with AlphaFold2. Representative  MSA subsets are selected by maximizing the diversity of their corresponding structures. For each representative MSA subset, we estimate its coevolutionary information using a Markov random field (MRF) model. Additional MSA subsets are constructed via enhanced sampling, which utilizes the differences in coevolutionary information embedded within the representative MSA subsets. 
The combined predictions are clustered with the representative structures extracted.
} \label{fig:SMICE_flowchart}
% \vskip -0.2in
\end{figure*}

SMICE consists of two key steps: the sampling step and the representative extraction step (Fig.\ref{fig:SMICE_flowchart}).

\subsection{Sampling step of SMICE}
SMICE embeds MSA sampling into generative probabilistic models and incorporates the coevolutionary information into the sampling criterion. 
A sequential sampling procedure is first applied to the full MSA to produce MSA subsets with diverse marginal statistics (amino acid proportions per residue), driven by a Bayesian framework.
When sampling each different MSA subset, the sequence sampling probability is computed under a Bayesian prior distribution of amino acid proportions. 
Varying the Bayesian prior distribution enables a broader exploration of the conformational space by sampling MSA subsets with distinct conservation patterns.
The sampled MSA subsets are used in AlphaFold2 to generate an initial set of structure predictions. 

Next, SMICE leverages coevolutionary information that would not have been captured in the marginal statistics used by sequential sampling. 
This begins with selecting the representative MSA subsets that predict the most structurally diverse conformations.
To utilize the differences in the coevolutionary information of these representative MSA subsets, a Markov Random Field (MRF) model \citep{kamisetty2013assessing} is fitted to each of the MSA subsets. We then rank sequences from the full MSA by their probability ratios under these competing MRF models. By selecting sequences that strongly favor one MRF model over another, we construct new MSA subsets enriched with specific coevolutionary information.
This enhanced (coevolution-aware) sampling is iterated for two cycles to ensure thorough exploration of the conformational space. 
The predicted structures from both the sequential sampling and the enhanced sampling are combined as the sampling result of SMICE.

\subsection{Representative extraction step of SMICE}
The representative extraction procedure is designed as follows: First, low-quality predictions are filtered out based on the pLDDT scores. 
Then, the variance of the residue contact map is calculated across the remaining structures, and the variable region of the protein is identified as a contiguous region that meets the following criteria: it must exhibit high variance either in its intra-region contact distances or in its inter-region contact distances (i.e., its contacts with the rest of the protein), while the contact distances within the rest of the protein remain stable.

Next, we cluster the high-quality structures based on their structural similarity in the variable region. After identifying the clusters and excluding the outliers, the structure with the highest pLDDT score within each cluster forms the final set of representative structures.

\section{Implementation Details}
\subsection{Hyperparameter configuration of random forest classifier}\label{supp:RF_param}

We used the R package \textit{ranger} (version 0.17.0) \citep{wright2017ranger} for training the random forest model.
The random forest classifier was optimized through 5-fold cross-validation. 
Each tree of the random forest classifier is trained with a bootstrap sample sampled from the training dataset.
Let $n_{\text{single-fold }}$ and $n_{\text{metamorphic}} $ represent the number of single-fold proteins and metamorphic proteins in the dataset, respectively. Let $n_{\text{total}} = n_{\text{single-fold }} + n_{\text{metamorphic}} $.
To achieve relatively balanced accuracy, the class weights for the single-fold and metamorphic protein classes are computed by $$\left(\frac{n_{\text{total}}}{2(1+\alpha)\cdot n_{\text{single-fold }} }, \frac{\alpha \cdot n_{\text{total}}}{2(1+\alpha)\cdot n_{\text{metamorphic}} } \right)$$
where $\alpha$ is a tuning parameter to address both class imbalance and the lower accuracy we empirically observed in the metamorphic protein class.

The hyperparameter search space is summarized in Table~\ref{tab:hyperparameters}.

\begin{table}[hbp]
\centering
\caption{Hyperparameter configuration for random forest classifier optimization}
\label{tab:hyperparameters}
\begin{tabular}{p{6cm}p{6cm}}
\toprule
\textbf{Hyperparameter} & \textbf{Search Space} \\
\midrule
Number of estimators & 500 \\
Maximum tree depth  & 5,10,15 \\
Minimum samples per leaf  & 6,8,10,12 \\
Class weight tuning parameter $\alpha$ & $1,2,4,6,8,10$ \\
Minimum impurity decrease & 0, 0.01 \\
\bottomrule
\end{tabular}
\end{table}

\end{document}